\documentclass[prx,twocolumn,superscriptaddress,citeautoscript,showpacs,amsart,longbibliography]{revtex4-1}
\usepackage{blindtext}
\usepackage{graphicx}
\usepackage{calrsfs}
\DeclareMathAlphabet{\pazocal}{OMS}{zplm}{m}{n}

\begin{document}
\title{$B_{\rm 1g}$ phonon anomaly driven by Fermi surface instability at intermediate temperature in YBa$_2$Cu$_3$O$_{7-\delta}$}
\author{Dongjin Oh}
\author{Dongjoon Song}
\author{Younsik Kim}
\affiliation{Center for Correlated Electron Systems, Institute for Basic Science, Seoul, 08826, Korea}
\affiliation{Department of Physics and Astronomy, Seoul National University, Seoul, 08826, Korea}

\author{Shigeki Miyasaka}
\author{Setsuko Tajima}
\affiliation{Department of Physics, Osaka University, Osaka 560-0043, Japan}

\author{Yunkyu Bang}
\affiliation{Asia Pacific Center for Theoretical Physics, Pohang 37673, Korea}

\author{Seung Ryong Park}
\email[Electronic address:$~$]{AbePark@inu.ac.kr}
\affiliation{Department of Physics, Incheon National University, Incheon 22012, Republic of Korea}

\author{Changyoung Kim}
\email[Electronic address:$~$]{changyoung@snu.ac.kr}
\affiliation{Center for Correlated Electron Systems, Institute for Basic Science, Seoul, 08826, Korea}
\affiliation{Department of Physics and Astronomy, Seoul National University, Seoul, 08826, Korea}

\begin{abstract}
We performed temperature- and doping-dependent high-resolution Raman spectroscopy experiments on YBa$_2$Cu$_3$O$_{7-\delta}$ to study $B$$_{\rm 1g}$ phonons. The temperature dependence of the real part of the phonon self-energy shows a distinct kink at $T=T_{\rm B1g}$ above $T$$_{\rm c}$ due to softening, in addition to the one due to the onset of the superconductivity. $T$$_{\rm B1g}$ is clearly different from the pseudogap temperature with a maximum in the underdoped region. The region between $T$$_{\rm B1g}$ and $T$$_{\rm c}$ resembles that of superconducting fluctuation or charge density wave order. While the true origin of the $B$$_{\rm 1g}$ phonon softening is not known, we can attribute it to a gap on the Fermi surface due to an electronic order. Our results may reveal the role of the $B$$_{\rm 1g}$ phonon not only in the superconducting state but also in the intertwined orders in multilayer copper oxide high-$T$$_{\rm c}$ superconductors.
\end{abstract}

\maketitle

In strongly correlated electron systems, instabilities with complex order parameters are often accompanied by electronic orders that cause various broken symmetries and Fermi surface instabilities in the ground state. The instabilities generically lead to an energy gap at the Fermi surface to lower the energy of the system. The most representative cases are Cooper pair and Peierls instabilities which are associated with superconductivity and charge density wave (CDW), respectively. In the former case, the superconducting ground state is created through coherent superposition of Cooper pairs and the superconducting gap is opened in the presence of a weak attractive interaction between electrons. Similarly in the latter case, coherent electron-hole pairs create CDW gap \cite{gruner2018density,li2019putting}. 

The gap opening driven by an electronic instability affects not only the electronic but also phononic excitation spectra. For example, it is well known that the $B$$_{\rm 1g}$ oxygen bond-buckling phonon in copper oxide high-$T$$_{\rm c}$ superconductors shows a drastic softening when the temperature is lowered below $T$$_{\rm c}$ \cite{altendorf1993temperature,limonov2000superconductivity,hewitt2004hole,bakr2009electronic}. The origin of the phonon anomaly was understood to be from a phonon self-energy effect caused by the opening of the superconducting gap (2$\Delta$); the phonon peak in the Raman spectrum exhibits narrowing due to the reduced scattering rate upon gap opening. In contrast, phonons with $\omega$$ > $2$\Delta$ show broadening since, for such phonons, scattering channels are available and their scattering rate increases due to the increased quasiparticle density of states (DOS). In this way, we can indirectly infer the existence of a superconducting gap. In addition to the superconductivity, this explanation can be also applied to other electronic orders such as CDW which gives rise to an energy gap \cite{raichle2011highly}. 

Among the various phonon modes in copper oxide superconductors, only the $B$$_{\rm 1g}$ phonon shows a drastic renormalization in the Raman spectra as we mentioned above. This result drew much interest since it may reflect the intimate link between superconductivity and $B$$_{\rm 1g}$ phonon. Indeed, the coupling matrix between electrons and $B$$_{\rm 1g}$ phonon at the Brillouin zone center ($q$=0) possesses a $d$-wave nature, which is the symmetry of the superconductivity as well as CDW form factor in cuprates \cite{devereaux1995charge,opel1999physical,fujita2014direct,comin2015symmetry,hamidian2016atomic}. Such character allows the $B$$_{\rm 1g}$ phonon to strongly couple with the electrons near the antinode where the superconducting gap is the maximum. As a result, a strong renormalization in the electronic structure appears in the form of kink or peak-dip-hump feature near the anti-nodal region \cite{devereaux2004anisotropic,cuk2004coupling}. A recent theoretical work pointed out that the $B$$_{\rm 1g}$ phonon may induce charge order in underdoped copper oxide superconductors \cite{banerjee2020emergent,banerjee2020intrinsic}. 


In spite of the predicted close relationship between the $B$$_{\rm 1g}$ phonon and electronic orders, a direct evidence for the relationship has not been observed. Discovery of such evidence may allow us to unify the understanding of the novel multiple orders in the phase diagram of high $T$$_{\rm c}$ materials. In this Letter, we present results of comprehensive doping- and temperature-dependent high-resolution Raman spectroscopy studies on YBa$_2$Cu$_3$O$_{7-\delta}$. Owing to the very high-statistics data, our results clearly show that phonon softening occurs not only in the superconducting state, but also at a temperature higher than $T$$_{\rm c}$ which has not yet been reported. The softening is robust in the underdoped region but is not observed for the most overdoped sample. The onset temperature of the phonon anomaly, $T$$_{\rm B1g}$, is distinguished from the pseudogap temperature $T$$^{\rm *}$. It is rather close to the CDW  temperature $T$$_{\rm CDW}$ \cite{blanco2014resonant} or superconducting fluctuation temperature $T$$_{\rm pair}$ \cite{rullier2011high,li2010diamagnetism,tallon2011fluctuations}. Even though our experimental results cannot point to the origin of the phonon anomaly, it is clear that a Fermi surface instability of an electronic order causes the softening of the $B$$_{\rm 1g}$ phonon. Our results not only highlight the important role of $B$$_{\rm 1g}$ phonon in the intertwined orders but also shed light on the possibility of their unified understanding in copper oxide superconductors.


\begin{figure}[b]
\includegraphics[width=8cm]{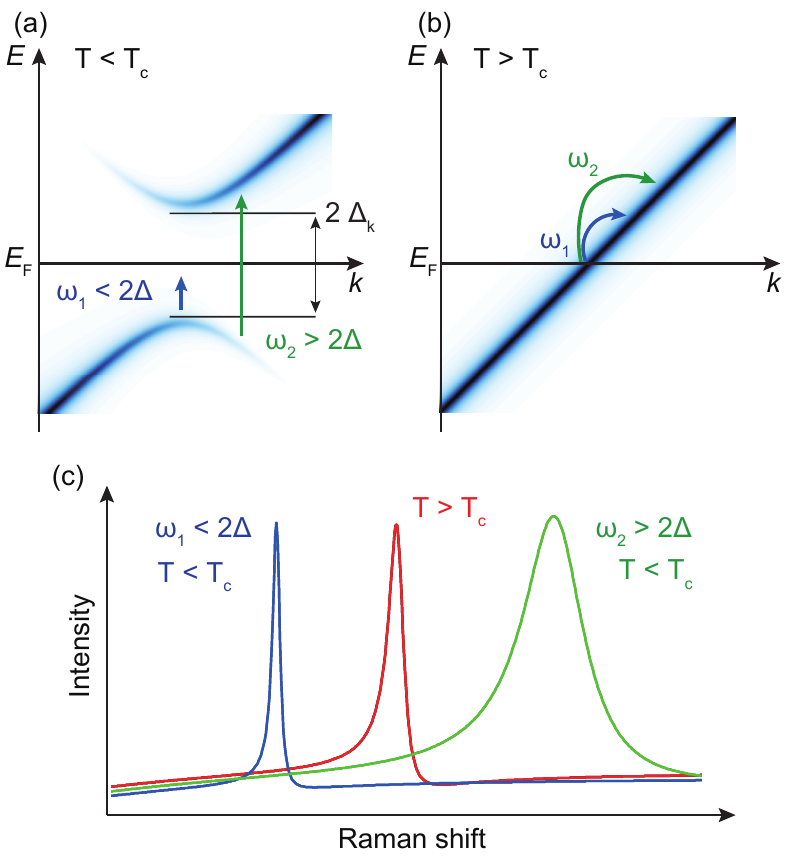}
\caption{Schematic of the low energy electronic structure of a superconductor (a) below and (b) above the critical temperature, $T$$_{\rm c}$. (c) Sketch of the phonon renormalization due to the phonon self-energy effect. The red curve represents a phononic Raman spectrum in the normal state. Blue and green curves correspond to the spectrum in the ordered state for the phonon energy $\omega < 2\Delta$ and $\omega > 2\Delta$, respectively.}
\end{figure}

\begin{figure}[htbp]
\includegraphics[width=8cm]{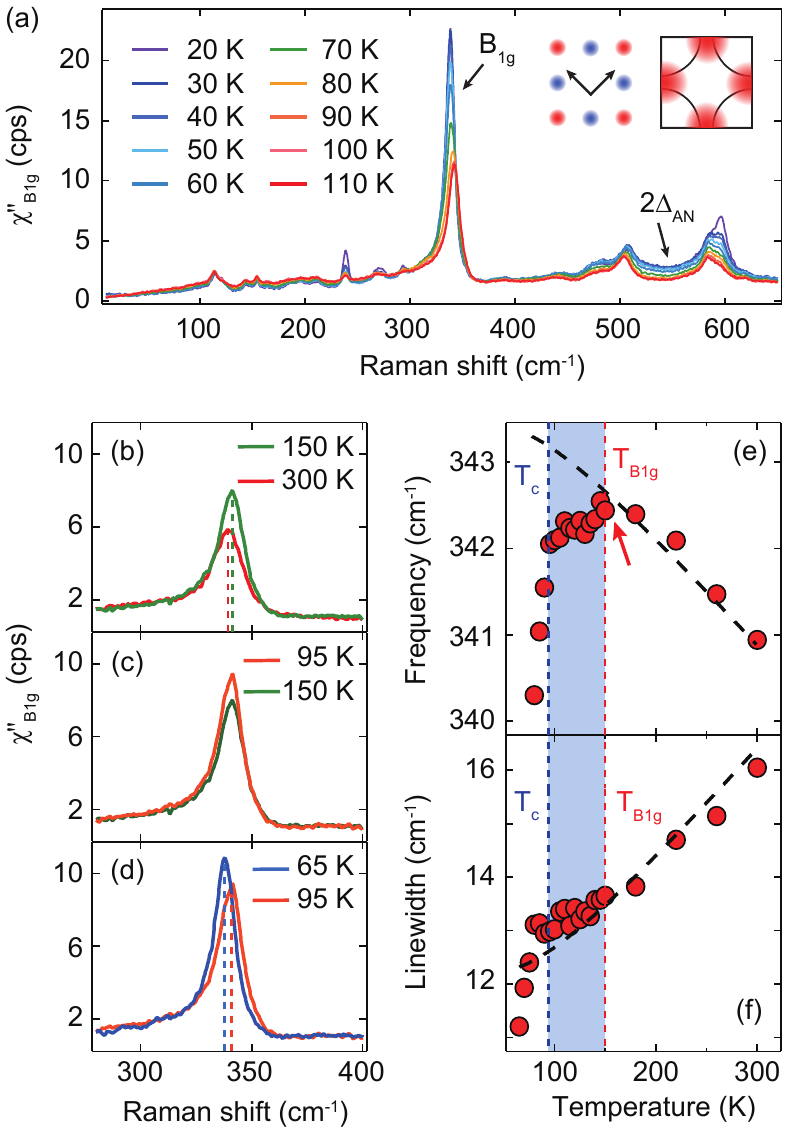}
\caption{(a) Temperature-dependent Raman response of optimally doped YBa$_2$Cu$_3$O$_{7-\delta}$  (OP94) single crystal below 110K. The peak at 340 cm$^{-1}$ corresponds to the $B$$_{\rm 1g}$  phonon and the broad peak around 550 cm$^{-1}$ is due to pair breaking peak. Compared are $B$$_{\rm 1g}$ phonon spectra measured at (b) 300 and 150 K, (c) 150 and 95 K, and (d) 95 and 65 K. Temperature-dependent frequency (e) and linewidth (f) extracted from a Fano line shape fitting. The black dashed curve corresponds to simulated frequency and linewidth of $B$$_{\rm 1g}$ phonon from the phonon anharmonicity model. The blue and red vertical dashed lines represent the superconducting transition temperature ($T$$_{\rm c}$) and onset temperature of the phonon anomaly ($T$$_{\rm B1g}$), respectively.}
\end{figure}

\begin{figure}[htbp]
\includegraphics[width=8cm]{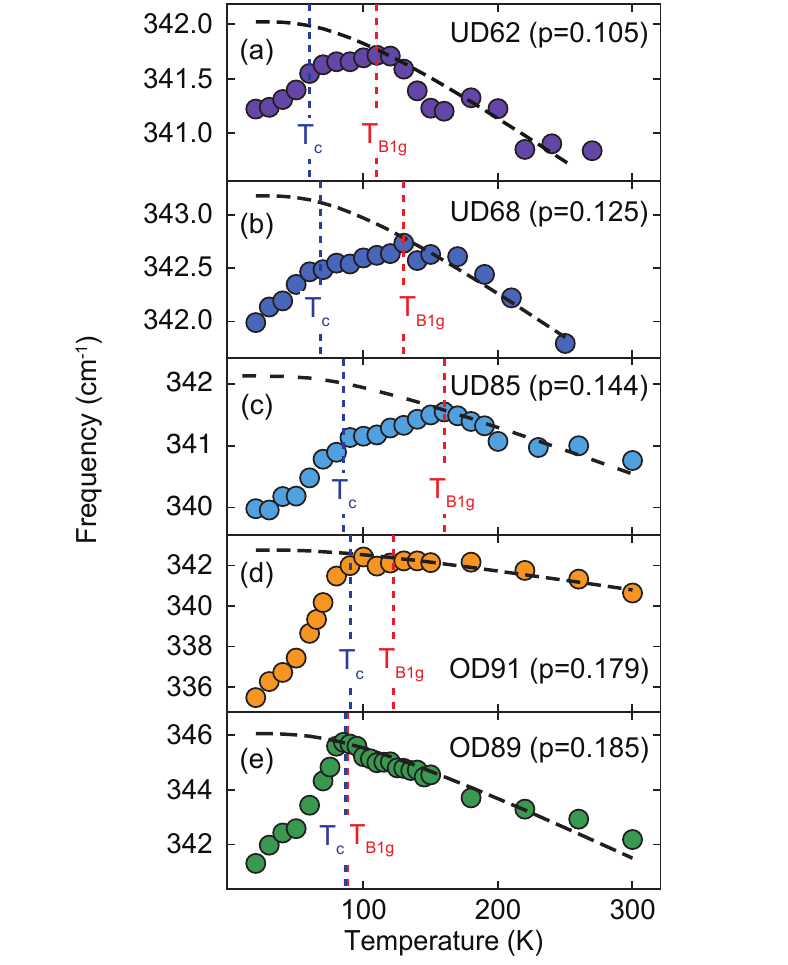}
\caption{Temperature dependent $B$$_{\rm 1g}$ phonon frequency from YBa$_2$Cu$_3$O$_{7-\delta}$ single crystals with different dopings. Results for UD62 (a), UD68 (b), UD85 (c), OD91 (d), OD89 (e). The hole doping concentration was determined using the method in Ref. \cite{liang2006evaluation}. $T$$_{\rm c}$ and ${T}$$_{\rm B1g}$ are indicated by blue and red vertical dashed lines, respectively.}
\end{figure}

\begin{figure}[h]
\includegraphics[width=8cm]{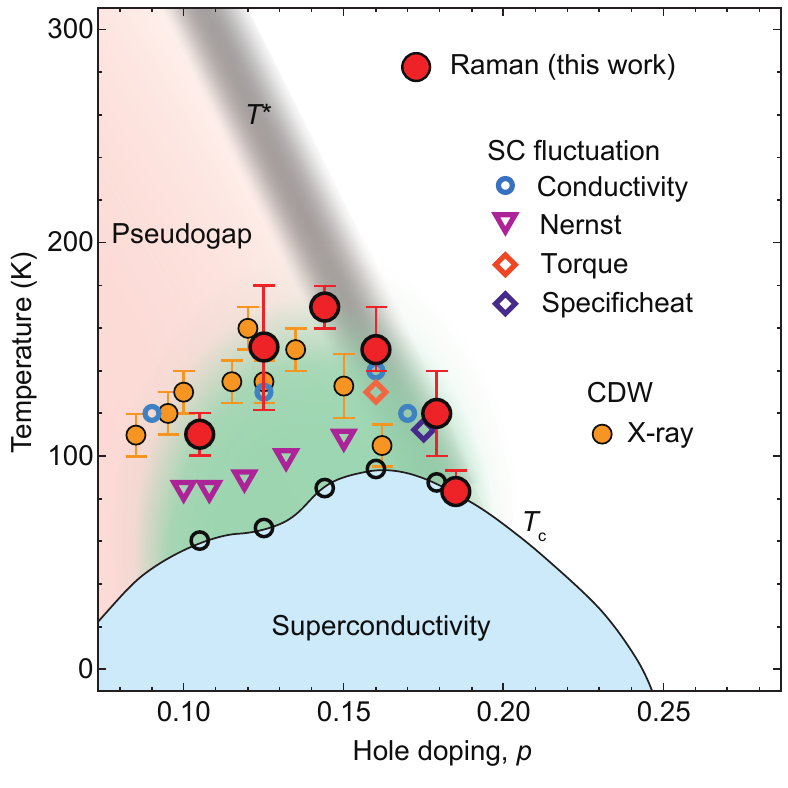}
\caption{YBa$_2$Cu$_3$O$_{7-\delta}$ phase diagram with various phases determined by various techniques. $T$$_{\rm B1g}$ extracted from our Raman spectroscopy data is plotted as red circles. Superconducting fluctuation temperature ($T$$_{\rm pair}$) in YBa$_2$Cu$_3$O$_{7-\delta}$  measured by various methods are also plotted: magnetoconductivity (blue circles) \cite{rullier2011high}, Nernst (violet inverted triangles) \cite{cyr2018pseudogap}, torque magnetometry (orange diamonds) \cite{li2010diamagnetism} and specific heat (purple diamonds) \cite{tallon2011fluctuations}. $T$$_{\rm CDW}$ measured by resonant x-ray scattering (RXS) \cite{blanco2014resonant} is plotted as yellow circles. Black shaded line represents the $T$$^*$ that can be found elsewhere \cite{zhao2017global,loret2019intimate}.} 
\end{figure}

We performed Raman spectroscopy with back-scattering geometry using 532 nm (2.3 eV) diode pumped solid state (DPSS) laser (Cobolt). The incident light was focused on the surface of the sample by a $\times$40 (N.A. 0.6) objective lens (Olympus). A spot size of the focused light was approximately 2 $\times$ 2 $\mu$m$^2$ and laser power was set below 0.6 mW to avoid laser-induced local heating. To reject the stray light close to 10 cm$^{-1}$, BragGrate notch filters (OptiGrate) were used. The superachromatic waveplate (Thorlabs) was used for polarization resolved Raman spectroscopy. The polarization of the incident and scattered light was set to be perpendicular to each other and parallel to the diagonal direction of the CuO$_2$ plaquette [Fig. 2(a), inset] to probe the signal from the $B$$_{\rm 1g}$ channel. The direction of the Cu$-$O bonding was determined by the angle-resolved polarized Raman scattering measurement. The Raman scattered light was collected by the same objective lens and collimated before entering the focusing lens in front of the spectrometer. The collected light was dispersed by a Jobin-Yvon Horiba iHR320 spectrometer with 1800 grooves/mm grating and detected by a thermoelectric cooled CCD. YBa$_2$Cu$_3$O$_{7-\delta}$ single crystals were mounted on an optical cryostat (Oxford) for temperature-dependent measurements. Optimally doped YBa$_2$Cu$_3$O$_{7-\delta}$ single crystals were annealed in a tube furnace with oxygen flow to control the oxygen content. The superconducting transition temperature was determined from the diamagnetic signal measured with a magnetic property measurement system (Quantum Design).

We first briefly touch upon how the gap in the electronic phase can affect the phonon peaks in Raman spectra. Plotted in Fig. 1 is a schematic of the low energy single particle spectral function across the superconducting gap and its effect on a phonon peak in Raman spectra. In the superconducting state, the superconducting gap opens at the Fermi surface while quasiparticle DOS at $\omega$ = $\pm$$\Delta$ is enhanced. It results in a singularity at $\omega$ = 2$\Delta$ in the real and imaginary parts of the phonon self-energy  \cite{zeyher1990superconductivity,nicol1993effect}. At $T<T_{\rm c}$, for a phonon with $\omega$$_{\rm ph}$ smaller (larger) than 2$\Delta$, softening (hardening) and narrowing (broadening) of the phonon is expected, in comparison to the $T >T_{\rm c}$ case [Fig. 1(c)]. This phenomenon can be intuitively understood as follows. In the normal state, all phonons can be scattered by coupling to the electrons near the Fermi level [Fig. 1(b)]. In the superconducting state, however, phonons with $\omega_{\rm ph} < 2\Delta$ cannot be scattered by electrons due to the lack of scattering channels. On the other hand, the scattering rate of $\omega_{\rm ph} > 2\Delta$ phonons increases due to the enhanced quasiparticle DOS [Fig. 1(a)]. 

Fig. 2(a) shows the $B$$_{\rm 1g}$ Raman response of optimally doped ($T$$_{\rm c}$ = 94 K) YBa$_2$Cu$_3$O$_{7-\delta}$. As the temperature decreases below $T$$_{\rm c}$, a broad electronic Raman continuum appears near 550 cm$^{-1}$. It is a signature of superconducting pair breaking, representing the superconducting gap near the antinodal region in the Brillouin zone [Fig. 2(a), inset] \cite{devereaux2007inelastic,devereaux1994electronic}. The strongest peak at 340 cm$^{-1}$ corresponds to the $B$$_{\rm 1g}$ phonon. Consistent with previous results, our results also show hardening in the normal state [Fig. 2(b)] and strong softening below $T$$_{\rm c}$ [Fig. 2(d)] \cite{altendorf1993temperature,limonov2000superconductivity,hewitt2004hole,bakr2009electronic}. However, an additional weak phonon softening was observed for a temperature range just above $T$$_{\rm c}$ [Fig. 2(c)]. To quantify the phonon softening, we extracted the frequency and linewidth of the $B$$_{\rm 1g}$ phonon by fitting the peak with a Fano line shape \cite{altendorf1993temperature,hewitt2004hole,bakr2009electronic,driza2012long}. Figs. 2(e) and (f) show the frequency and linewidth of the $B$$_{\rm 1g}$ phonon, respectively, as a function of temperature. The black dashed curves represent the simulated frequency and linewidth behavior using a phonon anharmonicity model \cite{altendorf1993temperature,bakr2009electronic}. Above 150 K, phonon hardening was well described by the phonon anharmonicity. The frequency of the $B$$_{\rm 1g}$ phonon begins to deviate from the anharmonic behavior around 150 K. The frequency decreases slowly until the temperature reaches $T$$_{\rm c}$, then decreases rapidly below $T$$_{\rm c}$. In Fig. 2(f), the linewidth of the $B$$_{\rm 1g}$ phonon also appears to deviate from the black dashed curve above $T$$_{\rm c}$. However, if higher order terms are added to the phonon anharmonicity model, the linewidth data can be fitted well from $T$$_{\rm c}$ to 300 K. Hence, we define the onset temperature of the phonon anomaly, dubbed as $T$$_{\rm B1g}$, from the temperature versus frequency data.

To verify the doping dependence of $T$$_{\rm B1g}$, we performed the same experiment and analysis on samples with various dopings. We find that not only the optimally doped sample but also underdoped samples (UD62, UD68 and UD85) [Fig. 3(a)-(c), respectively] clearly show the $B$$_{\rm 1g}$ phonon anomaly between $T$$_{\rm c}$ and $T$$_{\rm B1g}$. On the other hand, the phonon anomaly significantly weakens for slightly overdoped sample [Fig. 3(d)] and vanishes for the most overdoped sample [Fig. 3(e)]. This alludes to the fact that the instability that causes the $B$$_{\rm 1g}$ phonon anomaly disappears in the overdoped region.  

Plotted in Fig. 4 are the measured $T$$_{\rm B1g}$ (filled red circle) with the onset temperatures of various phases. The value of $T$$_{\rm B1g}$ is highest near slightly under doping ($p$=0.144) and decreases as the hole doping concentration increases or decreases. It is clearly seen that the doping dependence of $T$$_{\rm B1g}$ is quite different from that of $T$$^*$ because $T$$^*$ monotonously increases with decreasing hole doping concentration. This strongly suggests that the $B$$_{\rm 1g}$ phonon anomaly is not caused by the pseudogap. Instead, $T$$_{\rm B1g}$ appears to be close to the temperature scales of superconducting fluctuation, $T$$_{\rm pair}$. Except for $T$$_{\rm pair}$ measured by Nernst effect \cite{cyr2018pseudogap}, $T$$_{\rm pair}$ measured by magnetoconductivity \cite{rullier2011high}, torque magnetometry \cite{li2010diamagnetism} and specific heat \cite{tallon2011fluctuations} are quite close to $T$$_{\rm B1g}$. In addition, there is also a reasonable similarity between $T$$_{\rm CDW}$ and $T$$_{\rm B1g}$ \cite{blanco2014resonant}.

From the point of view that $T$$_{\rm B1g}$ corresponds to $T$$_{\rm pair}$, it can be understood that the phonon softening above $T$$_{\rm c}$ is caused by phonon self-energy effect due to formation of preformed Cooper pairs. Above $T$$_{\rm c}$, preformed Cooper pairs open a gap at the Fermi surface and enhance the quasiparticle DOS at $\omega$ = $\pm$$\Delta$ which has been clearly observed in a recent angle-resolved photoemission spectroscopy (ARPES) study on Bi2212 \cite{chen2019incoherent}. As a result, a singularity appears in the phonon self-energy which in turn causes the phonon softening as we mentioned earlier (Fig. 1). On the other hand, we also note that $T$$_{\rm B1g}$ is considerably higher than $T$$_{\rm pair}$ measured by Nernst effect \cite{cyr2018pseudogap}. The difference might be due to the uncertainty in the analysis to define $T$$_{\rm pair}$. Since the Nernst effect is very sensitive to formation of Cooper pairs, it is a powerful way to measure $T$$_{\rm pair}$. Nevertheless, there is a debate on how to determine the $T$$_{\rm pair}$ since the pseudogap also contributes to the Nenst signal. In the case of YBa$_2$Cu$_3$O$_{7-\delta}$, superconductivity and pseudogap contribute to a positive and negative Nernst signal, respectively \cite{cyr2018pseudogap}. The pseudogap contribution interferes with the signal from preformed Cooper pairs and makes it difficult to precisely define $T$$_{\rm pair}$. As a result, the $T$$_{\rm pair}$ measured by the Nernst effect underestimates the actual $T$$_{\rm pair}$ where Cooper pairs start forming.

Although $T$$_{\rm B1g}$ does not match previously reported $T$$_{\rm CDW}$ \cite{blanco2014resonant} very well, we cannot entirely rule out the possibility that CDW order is the origin of $B$$_{\rm 1g}$ phonon softening in the intermediate temperature range since the overall doping dependence of $T$$_{\rm B1g}$ is very similar to that of $T$$_{\rm CDW}$. CDW is the most common electronic order which softens phonon through the Kohn anomaly \cite{gruner2018density,le2014inelastic}. In this case, a giant phonon anomaly can occurs at $q$ = $q$$_{\rm CDW}$. However, $B$$_{\rm 1g}$ phonon softening in our results cannot be reconciled with the Kohn anomaly since Raman spectroscopy can probe only $q\sim0$ phonons. It is more natural to interpret that phonon softening comes from the phonon self-energy effect due to the Fermi surface instability induced by CDW \cite{raichle2011highly}. Based on these considerations, the mismatch between $T$$_{\rm CDW}$ and $T$$_{\rm B1g}$ can be understood as follows. Although no CDW peak was observed in resonant X-ray scattering (RXS) studies, inelastic neutron scattering results showed the softening of $B$$_{\rm 1g}$ phonon at $q=q_{\rm CDW}$ even for the most overdoped YBa$_2$Ca$_3$O$_7$ \cite{raichle2011highly,blanco2014resonant}. It suggests existence of CDW fluctuation even for the doping range where no static CDW order is observed. If the dynamic CDW affects the phonon self-energy, $T$$_{\rm B1g}$ may exhibit a different temperature scale than that measured by RXS. This calls for further studies with various experimental tools in the doping range where CDW order exists. 

If the observed $B$$_{1g}$ phonon softening is indeed due to a CDW order, it is suggestive of an important role played by the $B$$_{\rm 1g}$ phonon in both the high temperature superconductivity and charge order. This notion is indeed supported by the result of a recent electronic Raman study in which the energy scale of 2$\Delta$$_{\rm CDW}$ was found to be the same as that of antinodal superconducting gap 2$\Delta$$_{\rm AN}$ \cite{loret2019intimate}. This result suggests the possibility that the $B$$_{\rm 1g}$ phonon might play a role (a secondary, if not the primary role) for the superconductivity and CDW orders. Our result may lend another evidence for the connection between the superconductivity and CDW. 

Regardless of its origin, it is likely that the $B$$_{\rm 1g}$ phonon softening occurs due to a gap at the Fermi surface. Both Cooper pairing and Peierls instabilities result in an energy gap at the Fermi surface in their single particle excitation spectra even though the pairing channels for superconductivity (electron-electron) and CDW (electron-hole) are different \cite{gruner2018density}. Since these two orders induce the Fermi surface gap, in particular, share the maximum gap near the antinodal region, both can induce the renormalization in the phonon self-energy and softening of the $B$$_{\rm 1g}$ phonon.

It is also worth noting that both electronic orders have a $d$-wave nature. The link between $d$-wave symmetry and $B$$_{\rm 1g}$ phonon has been considered for a long time \cite{devereaux1995charge,newns2007fluctuating} and has recently attracted renewed attention \cite{he2018rapid,banerjee2020emergent,hamidian2016atomic}. It is argued that the electron-phonon coupling in the presence of strong electronic correlation can create electronic orders such as CDW with a $d$-wave symmetry form factor and $d$-wave superconductivity. Within this picture, our experimental results may be viewed as an additional evidence that the $B$$_{\rm 1g}$ phonon plays an important role in the formation of electronic orders with a $d$-wave character.

The remaining question is why $B$$_{1g}$ phonon anomaly is not observed in the pseudogap phase. A possible explanation may come from the different nature of the pseudogap. Spectroscopic studies have shown that the pseudogap phase are not accompanied by a sharply depleted Fermi surface gapping (and thus no quasiparticle DOS enhancement at $\omega$ = $\pm$$\Delta$), which should result in a smooth change in the phonon self-energy. Such effect can be expected in the pair density wave (PDW) scenario, where the spatial modulation of the $d$-wave pairing should lead to an almost cancelled coherence factor \cite{agterberg2020physics}. Further investigations of the pseudogap and electronic order (CDW or preformed pairs) using complementary experimental techniques should lead to a comprehensive understanding of the novel phases.

We acknowledge the fruitful discussion with S. Jung and D. Wulferding. This work was supported by the Institute for Basic Science in Korea (Grant No. IBS-R009-G2). S. R. P. acknowledges support from the NRF (Grant No. 2020R1A2C1011439).

\bibliographystyle{apsrev4-1} 

\end{document}